\documentclass[12pt]{article}
\usepackage{graphicx}
\usepackage{cite}
\usepackage{vmargin}
\usepackage{amsmath}
\usepackage{amstext}
\usepackage{xspace}

\setmarginsrb{1.25in}{1.25in}{1.25in}{1.0in}{0.0in}{0.0in}{0.0in}{0.50in}

\newcommand\pubnumber{DPF2015-256}
\newcommand\pubdate{\today}

\def\princeton{Department of Physics, Brookhaven National Laboratory\\
Upton, NY 11973, USA}
\def\myemail{\footnote{Email:  mrmooney@bnl.gov}}

\def\Title#1{\begin{center} {\Large #1 } \end{center}}
\def\Author#1{\begin{center}{ \sc #1} \end{center}}
\def\Address#1{\begin{center}{ \it #1} \end{center}}

\newcommand\pubblock{\rightline{\begin{tabular}{l} \pubnumber\\
         \pubdate  \end{tabular}}}
\newenvironment{Abstract}{\begin{quotation}}{\end{quotation}}
\newenvironment{Presented}{\begin{quotation} \begin{center}
                         \end{center}
      \begin{center}\begin{large}}{\end{large}\end{center} \end{quotation}}

\begin{document}
\begin{titlepage}
\pubblock

\vfill
\Title{The MicroBooNE Experiment and the Impact of Space Charge Effects}
\vfill
\Author{ Michael Mooney\myemail \\ \vspace{1mm} \rm On behalf of the MicroBooNE Collaboration}
\Address{\princeton}
\vfill
\begin{Abstract}
MicroBooNE is an experiment designed to both probe neutrino physics phenomena and develop the LArTPC (Liquid Argon Time Projection Chamber) detector technology.  The MicroBooNE experiment, which began taking data this year, is the first large LArTPC detector in the U.S.  This experiment is the beginning of a path of detectors (both on the surface and underground) envisioned for the U.S. SBL (Short-BaseLine) and LBL (Long-BaseLine) programs.  In order to interpret the data from the experiments on the surface, the impact of space charge effects must be simulated and calibrated.  The space charge effect is the build-up of slow-moving positive ions in a detector due to, for instance, ionization from cosmic rays, leading to a distortion of the electric field within the detector.  This effect leads to a displacement in the reconstructed position of signal ionization electrons in LArTPC detectors.  The LArTPC utilized in the MicroBooNE experiment is expected to be modestly impacted from the space charge effect, with the electric field magnitude changing by roughly 5\% (at a drift field of 500 V/cm) in some locations within the TPC.  We discuss the simulation of the space charge effect at MicroBooNE as well as calibration techniques that make use of a UV laser system and cosmic muon events.  A successful calibration of the space charge effect is imperative both to the success of the MicroBooNE physics program as well as to the development of LArTPC technology for future experiments.
\end{Abstract}
\vfill
\vspace{0.25in}
\begin{center}PRESENTED AT\end{center}
\begin{Presented}
DPF 2015\\
The Meeting of the American Physical Society\\
Division of Particles and Fields\\
Ann Arbor, Michigan, August 4--8, 2015\\
\end{Presented}
\vfill
\end{titlepage}

\section{MicroBooNE and LArTPC Technology}
\label{Introduction}

MicroBooNE (Micro Booster Neutrino Experiment) \cite{ref1,ref2} is a short-baseline (approximately 470 meters from source) neutrino experiment located at Fermilab in Batavia, Illinois, that utilizes the LArTPC (Liquid Argon Time Projection Chamber) technology \cite{ref3,ref4}.  The principal goal of this experiment is to determine the nature of the low-energy excess in electron (anti-)neutrinos observed at MiniBooNE (Mini Booster Neutrino Experiment).  MicroBooNE receives muon neutrinos from both the BNB (Booster Neutrino Beam) and NuMI (Neutrinos at the Main Injector) beamlines at Fermilab and recently began taking data while receiving neutrinos from both beamlines.

The experiment makes use of a cryostat filled with 170 tons of liquid argon, 89 tons of which is within the active volume of the TPC.  Argon was chosen for a variety of reasons: it is dense, abundant, ionizes easily, has a high electron lifetime, produces a large amount of scintillation light, and is transparent to its own scintillation light.  This makes argon a cost-effective choice in enabling a rich physics program that is driven by the combination of ionization electron signals and scintillation light in identifying particles from neutrino interactions.

\begin{figure}
\centering
\includegraphics[width=.50\textwidth]{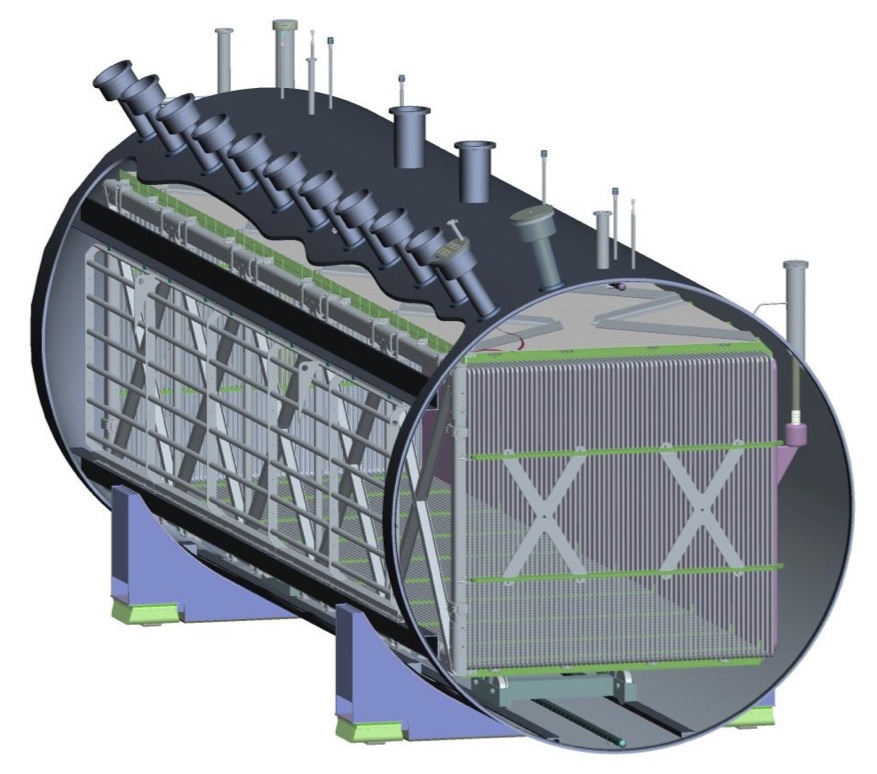}
\caption{Illustration of the MicroBooNE cryostat.  The anode plane of the TPC (which is 10.3 m long, 2.3 m tall, and 2.5 m wide) is located on the left side of the illustration, and the field cage can be seen at the downstream end of the cryostat.} 
\label{fig1}
\end{figure}

\begin{figure}
\centering
\includegraphics[width=.99\textwidth]{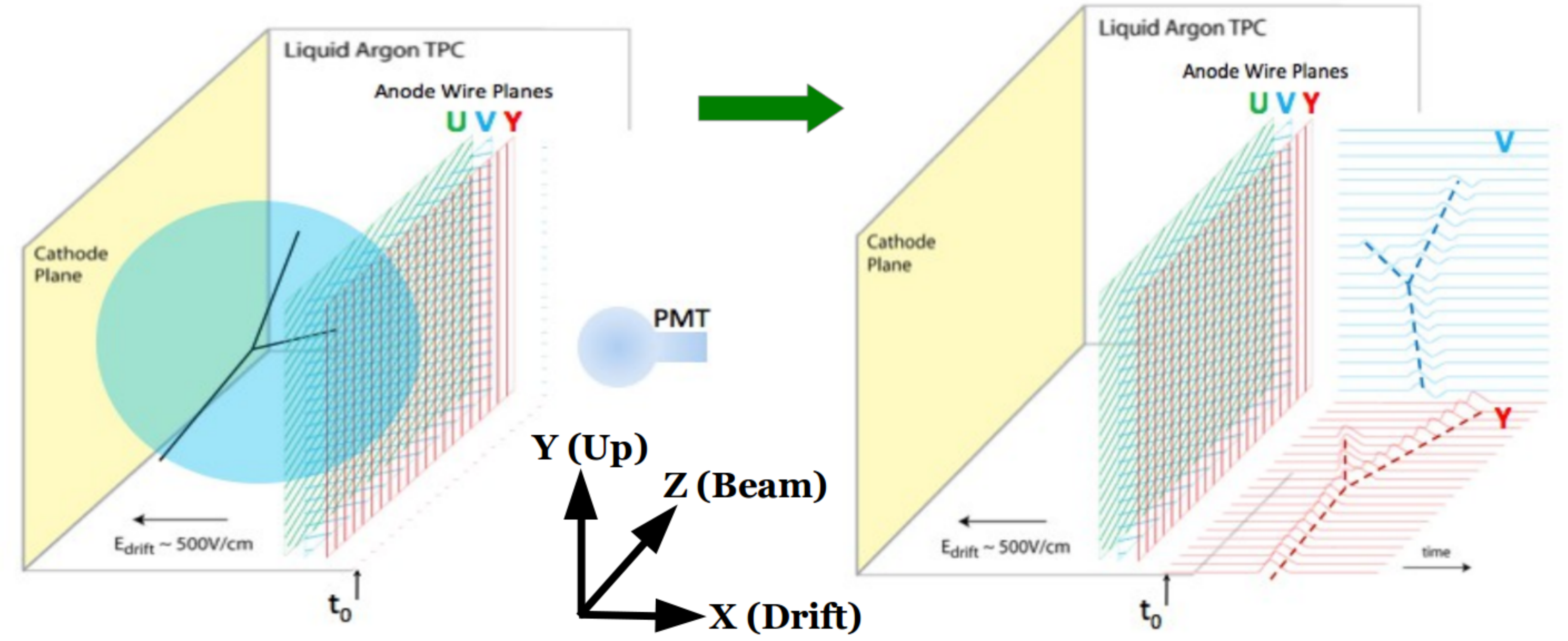}
\caption{Cartoon detailing how the LArTPC technology functions \cite{ref4}.  A neutrino interaction event in the bulk of the TPC results in the creation of ionization electrons which then drift toward three anode wire planes.  The signals induced on the wire planes, along with scintillation light collected by PMT's, allows for the reconstruction of three-dimensional particle trajectories in the detector.  Also shown is the coordinate system utilized by the space charge effect simulations described in Section \ref{Simulation}.} 
\label{fig2}
\end{figure}

Ionization electrons created in the argon due to impinging particles (including both cosmic muons and particles produced from neutrino interactions within the detector) are detected by means of three wire planes located at anode of the TPC.  The MicroBooNE cryostat and TPC is illustrated in Figure \ref{fig1}.  A drift electric field of 500 V/cm (design value) is maintined in the bulk of the TPC by means of a cathode held at -128 kV (design value) and 63 field cage tubes that line the sides of the TPC.  The electric field induced by this arrangement leads to the drifting of the ionization electrons toward the anode wire planes on the timescale of milliseconds.  The maximal drift distance of ionization electrons in the TPC is 2.56 m, the distance between the cathode and the anode.  Ionization electron clusters drift by the first two anode wire planes of the TPC and induce bipolar signals on the wires before being collected on the third wire plane (leading to a unipolar signal).  The ionization electron signals on the three wire planes, combined with timing information provided by PMT's (located behind the anode wire planes) via collection of scintillation photons, allows for the reconstruction of three-dimensional particle trajectories within the TPC bulk (see Figure \ref{fig2}).

\section{Space Charge Effect in LArTPC's}
\label{Simulation}

In order to correctly reconstruct the trajectories of particles that travel through the active volume of the TPC, it is essential to know very well the magnitude and direction of the drift electric field throughout the TPC bulk.  Nominally the electric field should be uniform throughout the TPC volume.  However, field effects such as the space charge effect may cause distortions in the electric field that result in distortions in the reconstructed position of ionization electron clusters detected by the TPC wire planes.  The space charge effect is the build-up of slow-moving positive ions in a detector due to, for instance, ionization from cosmic muons.  As MicroBooNE is an experiment on the surface with little overburden, the cosmic muon flux (10--20 muons per 4.8 ms readout window) is expected to create a significant amount of space charge (positive argon ions) that could modestly impact the drift electric field within the TPC active volume.

\begin{figure}
\includegraphics[width=.49\textwidth]{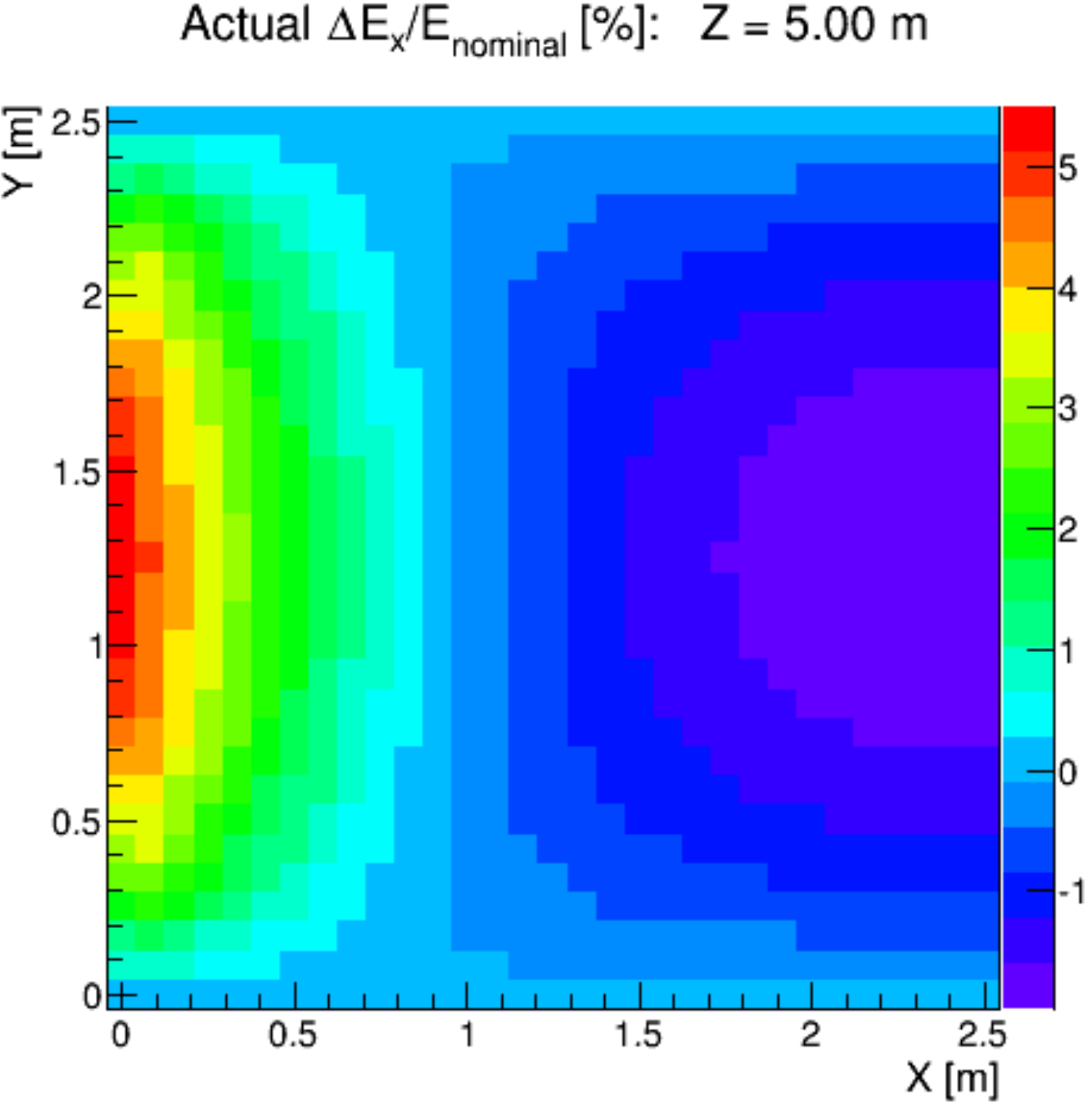}
\includegraphics[width=.49\textwidth]{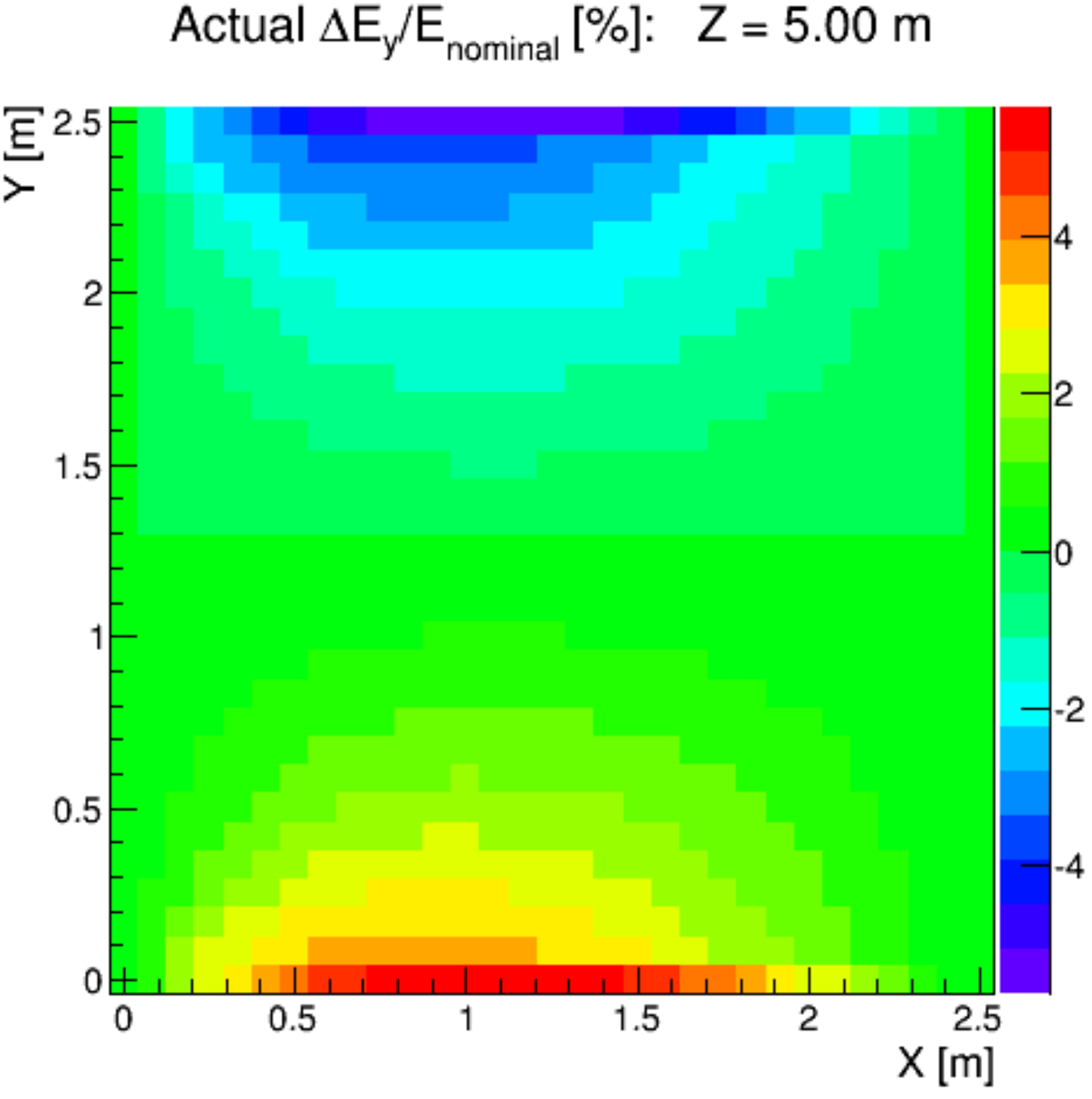}
\caption{Illustration of the effects of space charge on the drift electric field within the TPC at MicroBooNE (central slice in $z$, the coordinate aligned with the beam direction).  Shown is the variation of the electric field component in $x$, the drift direction (left) and $y$, the direction along the zenith axis (right), normalized to the nominal drift electric field magnitiude of 500 V/cm.} 
\label{fig3}
\end{figure}

We developed software to simulate the impact of space charge on the electric field within the TPC, along with the distortions in reconstructed ionization electron position at different points within the TPC bulk.  This simulation makes use of a Fourier series solution to the boundary value problem to solve for the electic field on a three-dimensional grid within the bulk of the TPC, an interpolation in between the grid points using radial basis functions to find the electric field everywhere in the TPC, and the RKF45 method for ray-tracing in order to simulate the distortions in reconstructed position of ionization electron clusters.  Some of the simulation results are shown in Figure \ref{fig3}, which illustrates the impact of space charge on the drift electric field.  At nominal drift field, the expected impact on the electric field is approximately 5\% in both the drift and transverse directions.

\begin{figure}
\includegraphics[width=.99\textwidth]{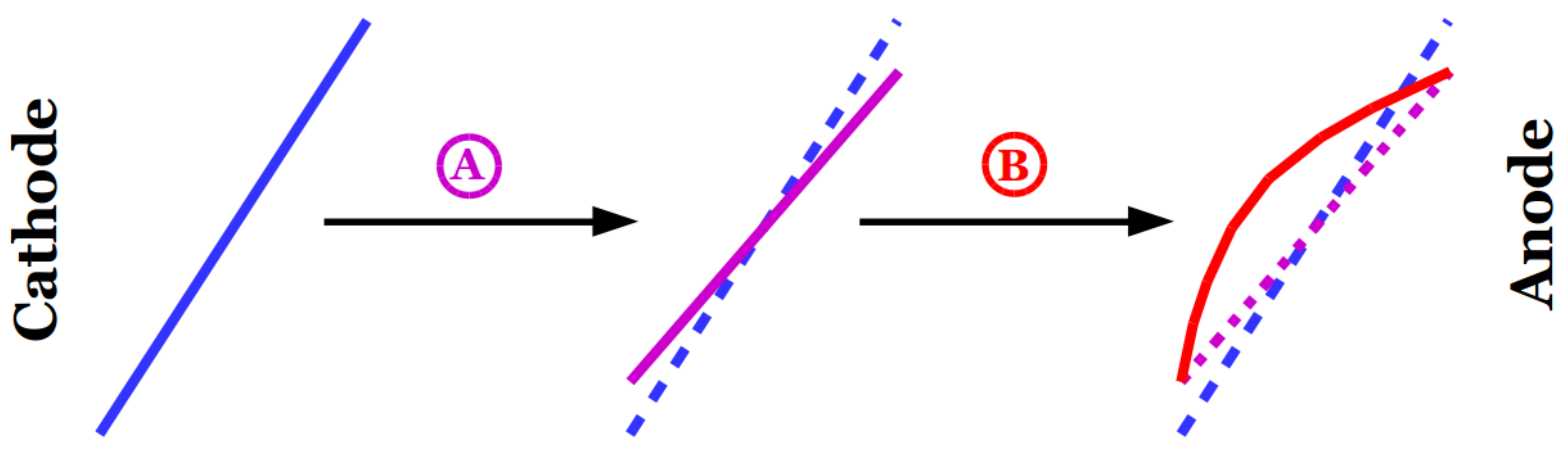}
\caption{Impact of the space charge effect on reconstructed tracks in the detector.  The impact on a reconstructed track can be broken up into two distinct features:  a squeezing of the sides of the tracks in the transverse TPC directions that can somewhat resemble a rotation (``A'') and a bowing of the track toward the cathode that is most pronounced in the middle of the TPC (``B'').} 
\label{fig4}
\end{figure}

In principle this effect could have modest impact in any TPC (liquid or gaseous) if the dimensions of the detector are large enough (in particular the maximal drift distance) to accumulate a high ion density.  In these cases, having a robust calibration method is necessary in order to account for the effect in particle trajectory reconstruction.  An example of the impact of the space charge effect on track reconstruction is shown in Figure \ref{fig4}.  We discuss a potential calibration method for the space charge effect at MicroBooNE in Section \ref{Calibration}.

\section{Calibration of Space Charge Effect}
\label{Calibration}

\begin{figure}
\includegraphics[width=.645\textwidth]{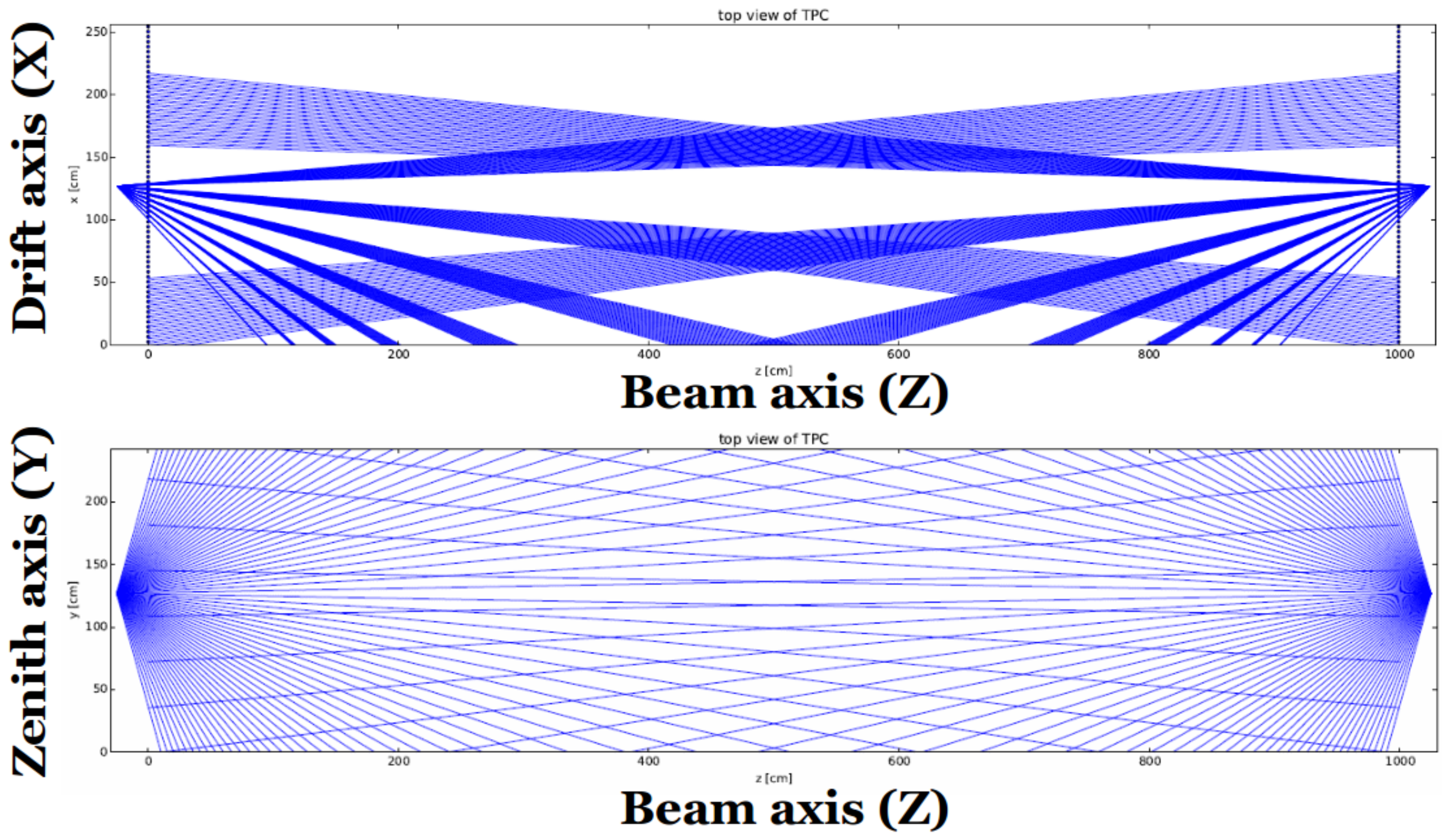}
\includegraphics[width=.345\textwidth]{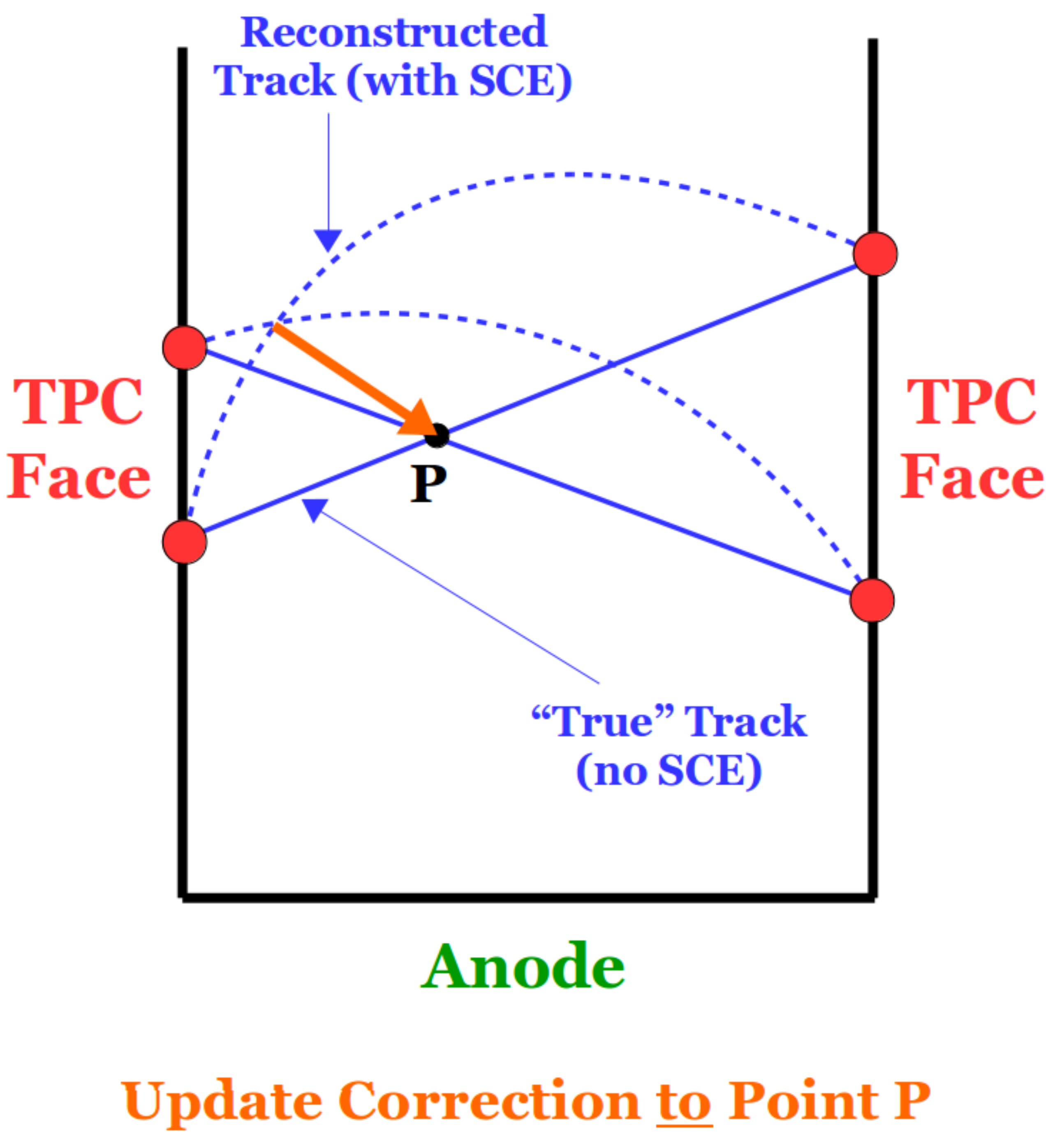}
\caption{One possible set of laser paths in the TPC from the upstream and downstream UV laser setups \cite{ref5} installed at MicroBooNE, shown from above (upper left) and from the side (lower left).  Also shown (right) is an illustration of the calibration scheme we utilize to correct the distortions in reconstructed position of the ionization clusters.  The near-crossing point of pairs of laser tracks (with known directions and start points) is compared against the near-crossing point of the reconstructed tracks, yielding a correction from the location of the reconstructed near-crossing point to the location of the ``truth'' near-crossing point.} 
\label{fig5}
\end{figure}

For the purpose of calibrating out field effects, two UV laser setups \cite{ref5} have been installed at MicroBooNE: one at each end of the TPC (in the beam direction).  One possible set of laser paths in the TPC provided by these two lasers is shown in Figure \ref{fig5}.  The laser beams must travel into the TPC through the field cage, which strongly limits the set of possible laser paths in the TPC active volume.

By combining one laser track from each of the two laser setups located at opposite ends of the TPC, we can estimate the distortion in reconstructed ionization electron cluster position at one point in the TPC.  This is possible because we know the true start points and directions (angles) for all of the laser tracks.  One calibration technique we have tried is taking the known crossing point of pairs of laser tracks, one from each end of the TPC, and comparing to the crossing point of the tracks reconstructed using the ionization electron signals produced in the TPC.  Since the laser pairs do not exactly intersect, we find the points-of-closest-approach of the two tracks (both for cases of either two ``truth'' tracks or two reconstructed tracks) and take the mid-point to serve as the ``near-crossing point'' of the two tracks.  If the truth tracks do not come closer than 2 cm to each other within the active volume of the TPC, the laser track pair is discarded from the calibration.  Furthermore, tracks that come closer to intersecting (again, using the truth information) are weighted more heavily in the calibration than those that don't come as close to each other.  For each pair of laser tracks, this calibration method provides a correction in reconstructed ionization electron position at one point in the TPC:  from the near-crossing point of the reconstructed tracks to the near-crossing point of the truth tracks.  An illustration of the calibration scheme is presented in Figure \ref{fig5}.

Note that this calibration scheme would also allow for the use of cosmic tracks in providing a correction to the space charge effect throughout the TPC.  Making use of cosmics in our calibration is important as it would allow for the filling in of the gaps between laser paths in the correction map.  These gaps are especially noticeable in the top view of Figure \ref{fig5}.  Since we don't know the exact trajectories of the cosmic muons in the TPC a priori, one would have to first obtain an estimate of the truth track by using an existing correction map (calculated by only using near-crossing points of two laser tracks) and applying these corrections to the reconstructed muon track.  The post-correction points from the reconstructed muon track can then be fit to provide an estimate of the true muon trajectory.  Then the method outlined above for two laser tracks can be used for a cosmic muon track and a laser track, or for two cosmic muon tracks.

\section{Results}
\label{Results}

\begin{figure}
\includegraphics[width=.49\textwidth]{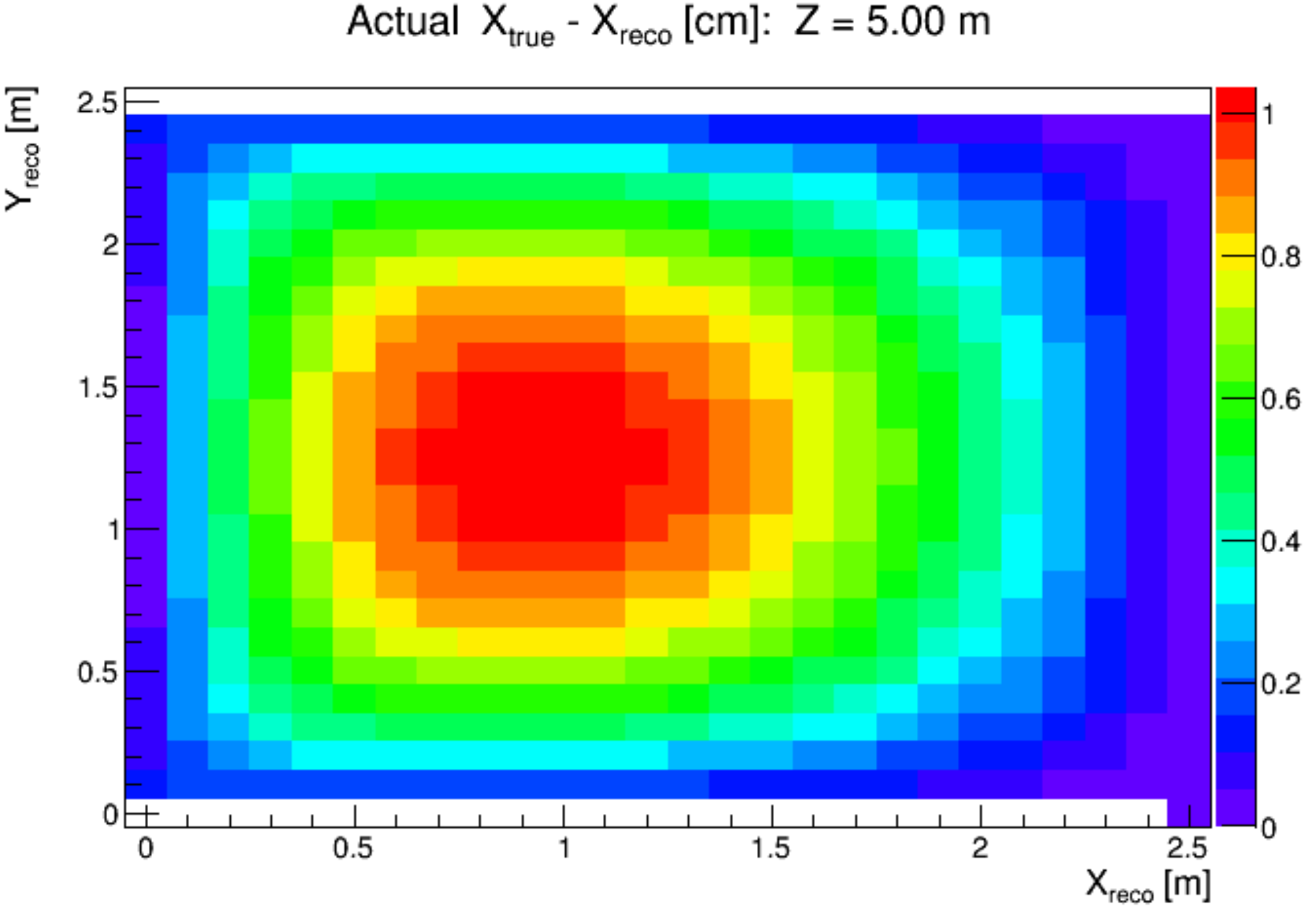}
\includegraphics[width=.49\textwidth]{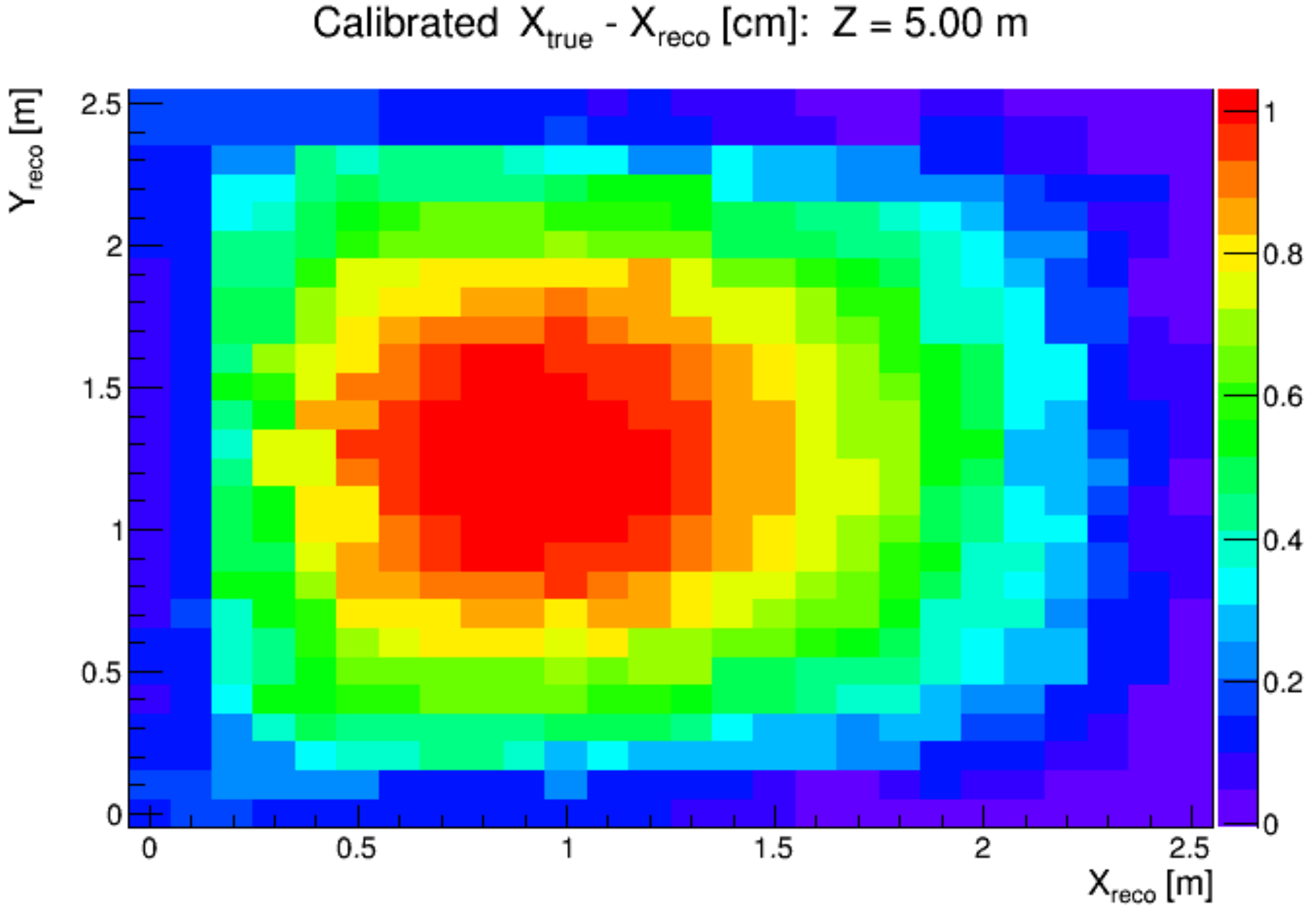}
\includegraphics[width=.49\textwidth]{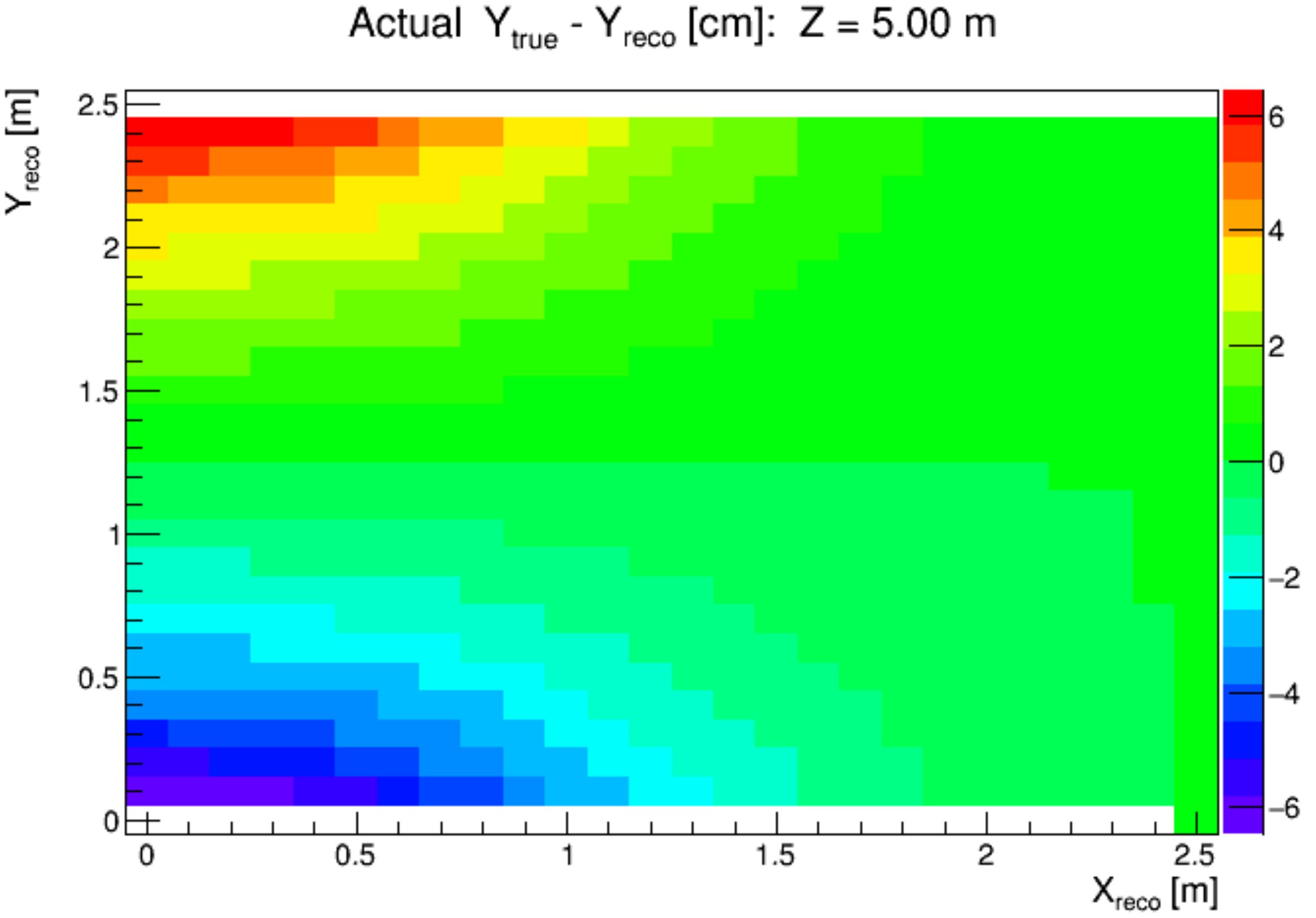}
\includegraphics[width=.49\textwidth]{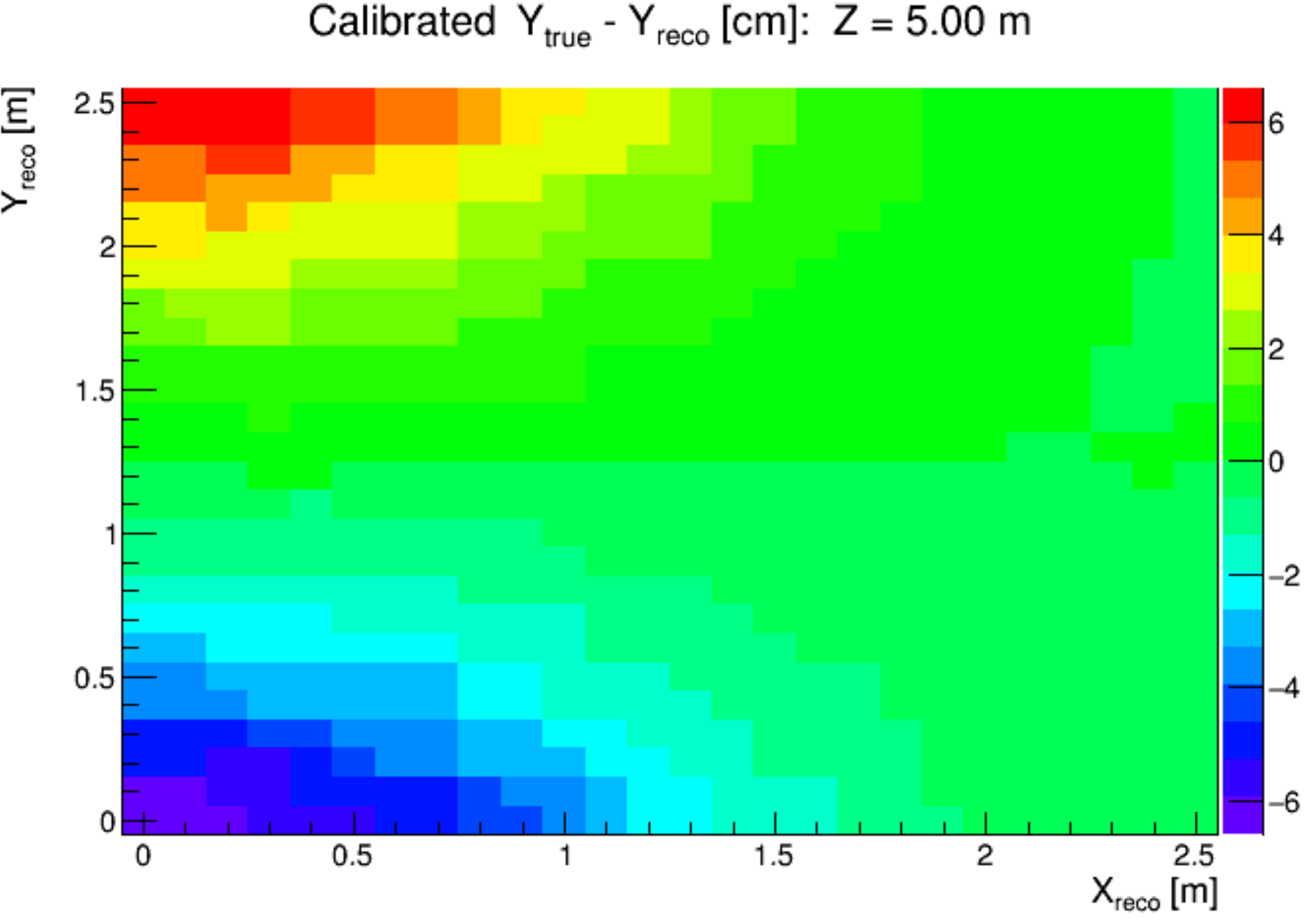}
\includegraphics[width=.50\textwidth]{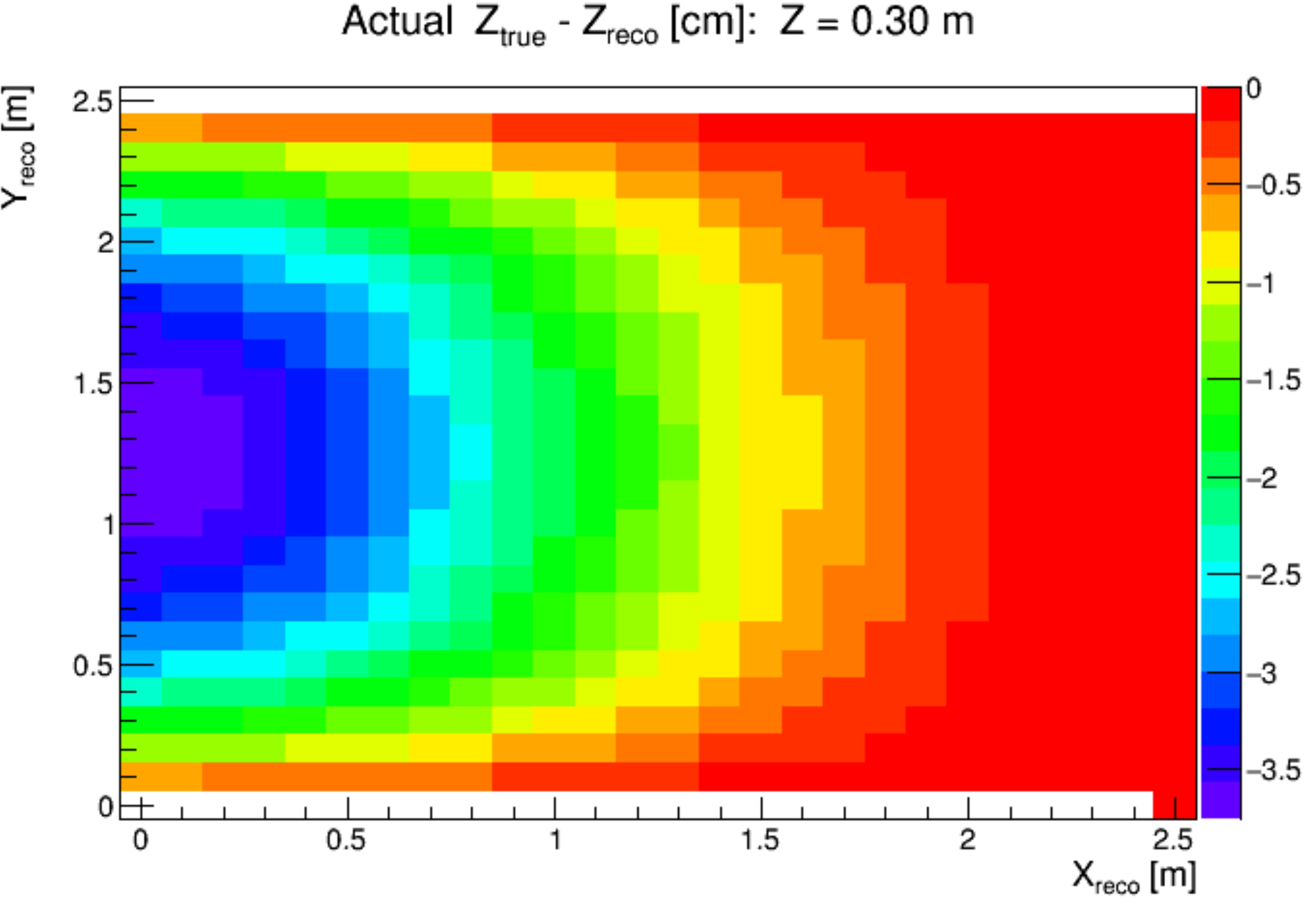}
\includegraphics[width=.50\textwidth]{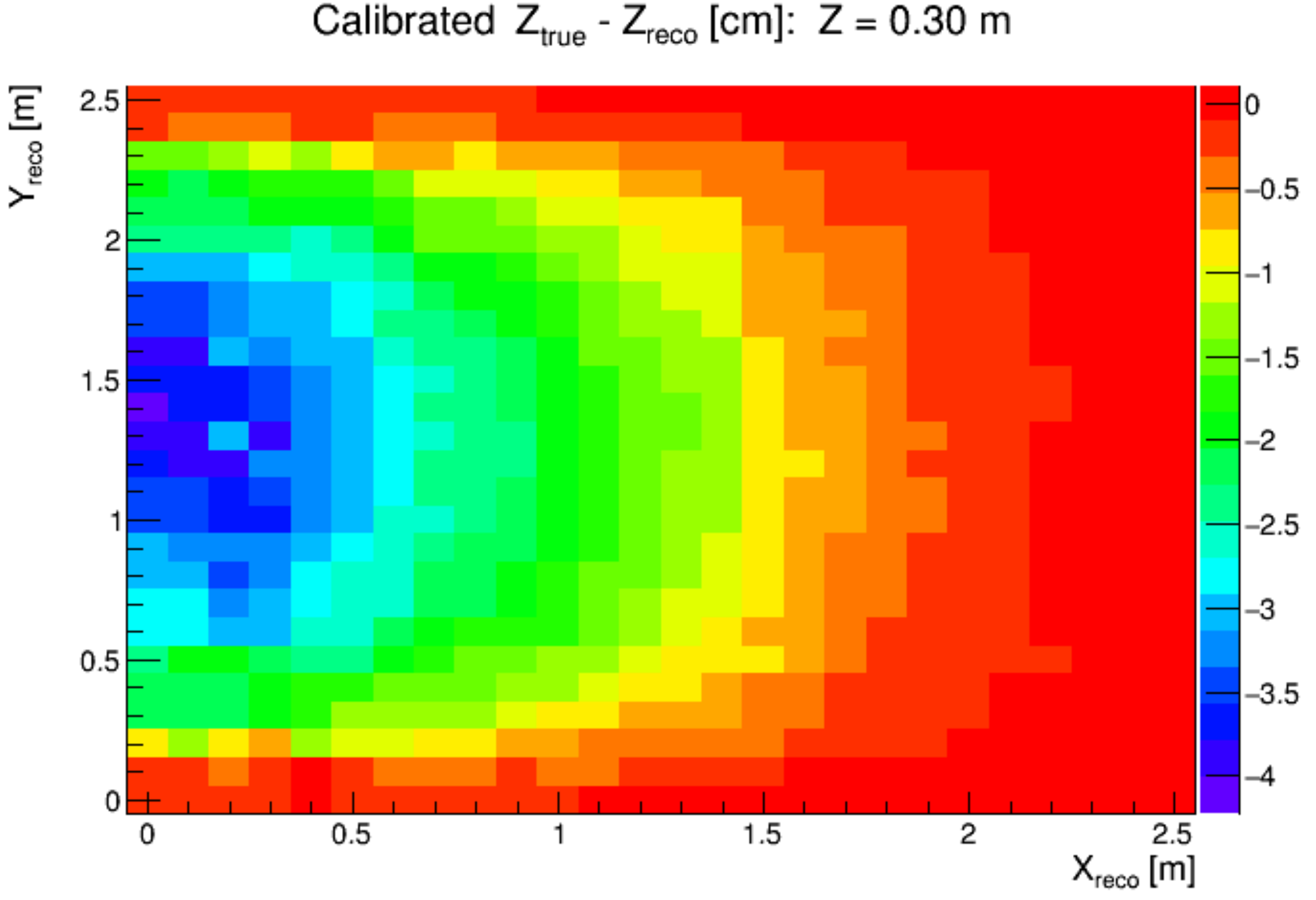}
\caption{Corrections to reconstructed ionization electron cluster position required to remove the space charge effect, both simulated (left column) and estimated from calibration (right column) by employing the calibration scheme discussed in Section \ref{Calibration}.  Results are shown in cm for corrections in the $x$ (top row), $y$ (middle row), and $z$ (bottom row) directions, and are plotted as a function of the position in the TPC at which the ionization electron clusters are reconstructed.  For the corrections in the $x$ and $y$ directions a central slice in $z$ is shown, while for the corrections in the $z$ direction a slice in $z$ closer to the end of the TPC ($z$ = 30 cm) is shown instead as the effect is bigger in this region of the TPC.} 
\label{fig6}
\end{figure}

Making use of the simulation tools described in Section \ref{Simulation} and the calibration technique outlined in Section \ref{Calibration} we are able to test the capability of our UV laser system in providing a correction to the space charge effect in the MicroBooNE TPC.  We take make use of a set of roughly 10,000 laser paths (shown in Figure \ref{fig5}) that are representative of what one laser ``scan'' might use.  In addition, we make use of 10,000 simulated cosmic muon tracks that are assumed to be high-momentum and thus approximately straight.  One could select such a sample of cosmic muons in actual data by making use of Multiple Coloumb Scattering (MCS) measurements.

Results are shown in Figure \ref{fig6}.  An interpolation is performed after the calibration to fill in parts of the TPC that did not see two truth tracks nearly cross.  The calibration method reconstructs quite well the true displacement of the ionization electron cluster position with respect to the location in the TPC that the ionization electron cluster originated from.  Note that in these results we are making a simplifying assumption that we know the true cosmic muon trajectory (as if it were a laser track), as opposed to estimating it using a fit to post-correction points of the reconstructed muon track, as described in the previous section.  This estimation will lead to further uncertainty in the correction obtained when making use of cosmic muon tracks as opposed to using only laser information.

\section{Conclusions}
\label{Conclusions}

For the LArTPC technology utilized by MicroBooNE to provide reliable reconstruction of particle trajectories in the detector, it is important for field effects such as the space charge effect to be properly accounted for.  Making use of a simulation package that we developed, we find the space charge effect at the nominal drift electric field of 500 V/cm to modify the electric field in the active volume of the TPC by roughly 5\% in both the drift and transverse directions.  UV laser setups at both ends of the TPC provide a means for the calibration of the space charge effect at MicroBooNE, while cosmic muons can provide additional coverage in the TPC across gaps between possible laser paths.  One calibration scheme, validated using the simulation software described above, has been shown to perform well in providing a correction map for the distortions in reconstructed ionization electron position due to the space charge effect.  However, more work must be done to provide a robust calibration of the space charge effect, including a more realistic estimation of the true trajectory of cosmic muons that are included in the calibration scheme.


\end{document}